\documentstyle[amssymb,12pt]{article}

\begin{document}

\title{NON-AUTONOMOUS DEGENERATE KdV SYSTEMS}
\author{ Ref\mbox{}ik Turhan  \\
{\small Department of Physics, Faculty of Arts and  Sciences}\\
{\small Middle East Technical University, 06531 Ankara-Turkey}}
\begin{titlepage}
\maketitle

\begin{abstract}
Non-autonomous degenerate KdV systems in (1+1) dimensions are considered for 
integrability classification. Integrability of the systems is associated with 
the existence of a recursion operator. Some new non-autonomous degenerate two-component 
KdV systems are found.
\end{abstract}

\end{titlepage}

\section*{INTRODUCTION}
~~~~Systems of integrable nonlinear partial differential equations have been 
constituting one of the main research areas for about two decades. Recently, in 
\cite{GKR} we have classified non-autonomous $N$-conmponent Korteweg-de Vries ($N$-KdV) 
systems 
\begin{eqnarray*}
q^{i}_{t}=q^{i}_{xxx}+s^{i}_{j k}(t)q^{j}q^{k}_{x},\;\
s^{i}_{j k}=s^{i}_{k j},\;i,j,k=1,2,...,N, 
\end{eqnarray*}
where dependent variables are $q^{i}=q^{i}(x,t)$ and  
found that the class of systems
\begin{eqnarray*}
 q^{i}_{t}=q^{i}_{xxx}+\frac{1}{\sqrt{t}}\tilde s^{i}_{j k}q^{j}q^{k}_{x},
\end{eqnarray*}
where $\tilde s^{i}_{j k}$ are constants, is integrable with the recursion operator
\begin{eqnarray*}
R^{i}_{j}&=& t \delta^{i}_{j}\,D^{2} +{2 \over 3}\,\sqrt{t}
\, \tilde s^{i}_{jk}\,q^{k}+ {1 \over 3}\delta^{i}_{j}x
+(\frac{1}{3}\,\sqrt{t}\, \tilde s^{i}_{jk}\,q^{k}_{x}+\frac{1}{6}
  \delta^{i}_{j}) \,D^{-1} \nonumber \\
&&  + \frac{1}{9} F^{i}_{\,\,lkj}\,q^{l}\,D^{-1}\,q^{k}\,D^{-1},
\end{eqnarray*}
where $D$ is the total $x$-derivative and $D^{-1}$ is the --formal-- inverse operator
($DD^{-1}=D^{-1}D=1$), provided that the identities
\begin{equation}
\tilde s^{k}_{p\,r} F^{i\,\,\,}_{ljk}+ \tilde s^{k}_{j\,r}
 F^{i\,\,\,}_{lpk}
+ \tilde s^{k}_{j\,p} F^{i\,\,\,}_{lrk}=0, \label{a1}
\end{equation}
where 
\begin{eqnarray*}
 F^{i\,\,\,}_{plj}=  \tilde s^{i}_{j\,k} \tilde s^{k}_{l\,p}
- \tilde s^{i}_{l\,k} \tilde s^{k}_{j\,p} ,
\end{eqnarray*}
are satisfied. Following Svinolupov's discovery that constants $\tilde s^{i}_{j k}$ satisfying (\ref{a1}) are associated with structure constants of  (commutative, non-associative) Jordan algebras \cite{SV2,SV4}, we called these systems as Non-autonomous Svinolupov Jordan KdV Systems. Moreover, we have also shown that systems associated with a certain type of Jordan algebras are not transformable to autonomous ones.  

In the present work, we consider a system of $N$ equations
\begin{equation}                                     	  
q^{i}_{t}=b^{i}_{j}q^{j}_{xxx}+s^{i}_{j k}(t)q^{j}q^{k}_{x}+y^{i}_{j}(t)q^{j},\;\
i,j,k=1,2,...,N,
\label{d2}                                           	  
\end{equation}                                       
and associate integrability of this system with the existence of a
recursion operator whose form we assume to be
\begin{equation}
R^{i}_{j}=z^{i}_{j}(t)D^{2}+a^{i}_{jk}(t)q^{k}\
+h^{i}_{j}(x,t)+(c^{i}_{jk}(t)q^{k}_{x}\
+w^{i}_{j}(t))D^{-1}.
\label{recop}
\end{equation}
Here, in (\ref{d2}) and (\ref{recop}) except for $b^{i}_{j}$ which are assumed to 
be constants, all the coefficient terms are introduced with their 
presumed dependence on the independent variables $x$ and $t$. Sufficient 
differentiability of these coefficients with respect to the independent variables 
is the other assumption made.

We first present the conditions due to the integrability criteria \cite{OLV}
\begin{equation}
R^{i}_{j,t}=K^{'i}_{r}R^{r}_{j}-R^{i}_{r}K^{'r}_{j},
\label{intcr}
\end{equation}
where $K^{'i}_{j}=b^{i}_{j}D^{3}+s^{i}_{k j}(t)q^{k}D+s^{i}_{j k}(t)q^{k}_{x}
+y^{i}_{j}(t)$ is the Fr\'{e}chet derivative of system (\ref{d2}).
 And then, we find solutions of these conditions for degenerate systems, 
i.e systems having $b^{i}_{j}$ such that $|b^{i}_{j}|=0$, in $N=2$.
 
\section*{INTEGRABILITY CONDITIONS}

~~~~The integrability criteria (\ref{intcr}) leads to some algebraic and differential conditions among the coefficient terms of the system (\ref{d2}) and the recursion operator 
(\ref{recop}). In the following proposition we present these conditions.

{\bf Proposition 1:} Let $q^{i}(x,t)$ be functions of $x$ and $t$ satisfying
equations (\ref{d2}) and admitting a recursion operator 
$R^{i}_{j}$ given in (\ref{recop}). Then the coefficient terms 
$b^{i}_{j}, s^{i}_{j k}(t), y^{i}_{j}(t), z^{i}_{j}(t), a^{i}_{jk}(t), h^{i}_{j}(x,t), 
c^{i}_{jk}(t), w^{i}_{j}(t)$ satisfy the following relations: 

\begin{displaymath}
\begin{array}{lr}
3b^{i}_{k}a^{k}_{jl}+b^{i}_{k}c^{k}_{jl}+s^{i}_{kl}z^{k}_{j}\
-2z^{i}_{k}s^{k}_{lj}-z^{i}_{k}s^{k}_{jl}-c^{i}_{kl}b^{k}_{j}=0,&\
s^{i}_{mk}a^{k}_{jl}-a^{i}_{km}s^{k}_{lj}=0,
\end{array}
\end{displaymath}
\begin{displaymath}
\begin{array}{lr}
3b^{i}_{k}a^{k}_{jl}+3b^{i}_{k}c^{k}_{jl}-z^{i}_{k}s^{k}_{lj}-2z^{i}_{k}s^{k}_{jl}=0,&\
a^{i}_{jk}b^{k}_{l}-b^{i}_{k}a^{k}_{jl}-3b^{i}_{k}c^{k}_{jl}+z^{i}_{k}s^{k}_{jl}=0,
\end{array}
\end{displaymath}
\begin{displaymath}
\begin{array}{lccr}
b^{i}_{k}z^{k}_{j}-z^{i}_{k}b^{k}_{j}=0,&\
b^{i}_{k}h^{k}_{j}-h^{i}_{k}b^{k}_{j}=0,&\
b^{i}_{k}c^{k}_{jl}-c^{i}_{jk}b^{k}_{l}=0,&\
w^{i}_{k}(s^{k}_{lj}-s^{k}_{jl})=0,
\end{array}
\end{displaymath}
\begin{displaymath}
\begin{array}{lcr}
c^{i}_{km}(s^{k}_{jl}-s^{k}_{lj})=0,&\
c^{i}_{jk}s^{k}_{ml}-s^{i}_{km}c^{k}_{jl}=0,&\
c^{i}_{jk}s^{k}_{ml}-s^{i}_{mk}c^{k}_{jl}=0,
\end{array}
\end{displaymath}
\begin{equation}
\begin{array}{c}
a^{i}_{jk}s^{k}_{ml}+a^{i}_{km}s^{k}_{jl}+c^{i}_{kl}s^{k}_{jm}\
-s^{i}_{mk}a^{k}_{jl} -s^{i}_{mk}c^{k}_{jl}-s^{i}_{kl}a^{k}_{jm}=0,
\end{array}
\label{inkos}
\end{equation}
\begin{displaymath}
\begin{array}{lcr}
b^{i}_{k}a^{k}_{jl}+s^{i}_{lk}z^{k}_{j}-z^{i}_{k}s^{k}_{lj}-a^{i}_{kl}b^{k}_{j}=0,&\
s^{i}_{lk}h^{k}_{j}-h^{i}_{k}s^{k}_{lj}=0,&\
s^{i}_{kl}h^{k}_{j}-h^{i}_{k}s^{k}_{jl}=0,
\end{array}
\end{displaymath}
\begin{displaymath}
\begin{array}{lr}
z^{i}_{j ,t}-3b^{i}_{k}h^{k}_{j,x}-b^{i}_{k}w^{k}_{j}\
-y^{i}_{k}z^{k}_{j}+z^{i}_{k}y^{k}_{j}+w^{i}_{k}b^{k}_{j}=0,&\
w^{i}_{j,t}-y^{i}_{k}w^{k}_{j}+w^{i}_{k}y^{k}_{j}=0,
\end{array}
\end{displaymath}
\begin{displaymath}
\begin{array}{lr}
a^{i}_{jl,t}+a^{i}_{jk}y^{k}_{l}-s^{i}_{lk}h^{k}_{j,x}-s^{i}_{lk}w^{k}_{j}\
-y^{i}_{k}a^{k}_{jl}+a^{i}_{kl}y^{k}_{j}+w^{i}_{k}s^{k}_{jl}=0,&\
b^{i}_{k}h^{k}_{j,2x}=0,
\end{array}
\end{displaymath}
\begin{displaymath}
\begin{array}{lr}
c^{i}_{jl,t}+c^{i}_{jk}y^{k}_{l}-s^{i}_{kl}w^{k}_{j}-y^{i}_{k}c^{k}_{jl}+c^{i}_{kl}y^{k}_{j}=0,&\
h^{i}_{j,t}-b^{i}_{k}h^{k}_{j,3x}-y^{i}_{k}h^{k}_{j}+h^{i}_{k}y^{k}_{j}=0.
\end{array}
\end{displaymath}
\noindent
This proposition is the straight forward result of calculating (\ref{intcr}) together 
with (\ref{d2}) and (\ref{recop}).

 Our basic aim is to determine the integrable systems (or classes of systems) with their associated recursion operators by solving the whole system (\ref{inkos}). For this purpose we base our classification on $b^{i}_{j}$ and solve conditions (\ref{inkos}) for each of specially chosen different forms of $b^{i}_{j}$ matrix. Having chosen a certain $b^{i}_{j}$, we first solve the algebraic part of the equations in (\ref{inkos}). Although the algebraic part is quite large even in $N=2$, because the constituent equations are polynomials of order at most two in the unknowns, they can be solved by computer algebra systems conveniently. We used REDUCE and MuPAD software for the computations. Each nontrivial solution to the algebraic part is then subjected to the remaining differential conditions.

  In general, for a certain $b^{i}_{j}$ matrix there are many solutions to the integrability conditions (\ref{inkos}). However,  some of these solutions give rise to systems which are not valuable: Decoupled systems, trivially coupled systems like $u_{t}=F[u,t],v_{t}=G[u,t]$ or completely linear systems. Besides these,  some solutions require the recursion operator to be (or proportional to) identity, in  case of which (\ref{intcr}) is obviously inconclusive for the integrability of the system at hand. These basically constitute our criteria for triviality of a solution to (\ref{inkos}). We discard all such trivial cases.

\section*{CLASSIFICATION} 

~~~~In this algorithm each different $b^{i}_{j}$ matrix cause a different set of 
solutions to (\ref{inkos}). We call a collection of systems obtained under a certain 
$b^{i}_{j}$ as a class and naturally identify each class with its $b^{i}_{j}$ matrix. Here, difference of matrices refers to dissimilarity of them. 

Since our particular concern is the degenerate 2-KdV systems in this work, we consider the following degenerate and dissimilar forms for  $b^{i}_{j}$  matrix:
\begin{equation}
b(1)=
\left(\begin{array}{cc}
1 & 0 \\
0 & 0
\end{array} \right),\;\;
b(2)=
\left(\begin{array}{cc}
0 & 1 \\
0 & 0
\end{array} \right).
\label{mtrcs}
\end{equation}

{\bf Remark 1:} Up to an overall factor, the above two matrix which are in Jordan canonical form constitute all the non-zero degenerate matrices in $N=2$ under the similarity transformations. I.e any non-zero, degenerate, $2\times 2$ matrix $b^{k}_{l}$ is transformable to one of the above two with an overall factor by an invertible (similarity) transformation $M^{i}_{j}$ as $\tilde{b}^{i}_{j}=M^{i}_{k}b^{k}_{l}(M^{-1})^{l}_{j}$, where $M^{i}_{j}$ are constants. Existence of such invertible $M^{i}_{j}$'s is assured by Jordan canonical form theorem of matrices \cite{HK}. The null matrix and the overall factors for $b(1)$ and $b(2)$ are irrelevant for our classification even if these factors are taken as functions of $t$. They can always be absorbed into the left hand side of evolution system (\ref{d2}) by redefining the evolution parameter $t$.

{\bf Remark 2:} The above mentioned similarity transformations correspond to change of dependent variables $\tilde{q}^{i}=M^{i}_{k}q^{k}$ and the integrability criteria (\ref{intcr}) is invariant under such transformations. We do not distinguish a system from its $M^{i}_{j}$ transformed form 
\begin{eqnarray*}
M^{i}_{k}q^{k}_{t}&=&M^{i}_{k}b^{k}_{l}(M^{-1})^{l}_{j}M^{j}_{r}q^{r}_{xxx}+\
M^{i}_{k}s^{k}_{jl}(M^{-1})^{j}_{r}(M^{-1})^{l}_{p}M^{r}_{s}q^{s}M^{p}_{\nu}q^{\nu}_{x}\\
&&+M^{i}_{k}y^{k}_{j}(M^{-1})^{j}_{p}M^{p}_{s}q^{s}.
\end{eqnarray*}
\noindent
Moreover, we suggest to use the form of a KdV system where its 
$b^{i}_{j}$ is put in Jordan canonical form, as the standard form for it.

\subsection*{a) Systems associated with $b^{i}_{j}= b(1)$ }
~~~~Although our main concern is the non-autonomous systems, solutions of integrability 
conditions (\ref{inkos}) with $b^{i}_{j}=b(1)$ in (\ref{mtrcs}) give rise to 
autonomous systems as well. In the following list we first give these  autonomous systems with their respective recursion operators.
\newline {\bf i)}
\begin{equation}
\begin{array}{c}
u_{t}=u_{xxx}+3c_{1}uu_{x}\\
v_{t}=c_{1}uv_{x}+c_{2}vu_{x}
\end{array},\,\,
R=
\left(
\begin{array}{cc}
D^{2}+2c_{1}u+c_{1}u_{x}D^{-1} & 0  \\
c_{2}v+c_{1}v_{x}D^{-1} & 0
\end{array}  \right),
\label{sys1}
\end{equation}
\noindent
where $c_{1}$ and $c_{2}$ are arbitrary constants \cite{GK1},\cite{FUCH}.
\newline {\bf ii)}
\begin{equation}
\begin{array}{c}
u_{t}=u_{xxx}+3c_{1}uu_{x}+c_{2}vv_{x}\\
v_{t}=c_{1}(uv)_{x}
\end{array},\,\,
R=
\left(
\begin{array}{cc}
D^{2}+2c_{1}u+c_{1}u_{x}D^{-1} & c_{2}v  \\
c_{1}(v+v_{x}D^{-1}) & 0
\end{array} \right),
\label{sys2}
\end{equation}
\noindent
where $c_{1}$ and $c_{2}$ are arbitrary constants. This system is equivalent to the Ito system via $u \rightarrow \frac{2}{c_{1}}u$, 
$v \rightarrow \frac{2}{\sqrt{c_{1}c_{2}}}v$ transformation \cite{ITO}.
\newline{\bf iii)}
\begin{equation}
\begin{array}{c}
u_{t}=u_{xxx}+3c_{1}uu_{x}\\
v_{t}=(c_{1}u+c_{2}v)v_{x}
\end{array},\,\,
R=
\left(
\begin{array}{cc}
D^{2}+2c_{1}u+c_{1}u_{x}D^{-1} & 0  \\
c_{1}v_{x}D^{-1} & c_{2}v
\end{array} \right),
\label{sys3}
\end{equation}
\noindent
where $c_{1}$ and $c_{2}$ are arbitrary constants. 

At this point, for the above list of systems we have the following statement.

{\bf Proposition 2:} Systems (\ref{sys1}), (\ref{sys2}) and (\ref{sys3}) constitute all the nontrivial autonomous systems in $b(1)$ class.

The followings are the non-autonomous systems with their respective recursion operators in $b(1)$ class.
\newline {\bf i)}
\begin{equation}
\begin{array}{c}
u_{t}=u_{xxx}+c_{1}t^{-(\alpha +2/3)}vv_{x}+c_{2}t^{-(\alpha +1)}v,\\
v_{t}=t^{-2/3}vv_{x},
\end{array}
\label{sys4}
\end{equation}
\noindent
where $c_{1}$, $c_{2}$ and $\alpha$ are arbitrary constants. 
This system admits the recursion operator
\begin{equation}
R=
\left(
\begin{array}{cc}
tD^{2}+\frac{x}{3}+\alpha D^{-1} & t^{-\alpha}\left(c_{1}t^{1/3}v+c_{2}D^{-1}\right)  \\
0 & t^{1/3}v+\frac{x}{3}
\end{array} \right).
\label{rec4}
\end{equation}
\newline {\bf ii)}
\begin{equation}
\begin{array}{c}
u_{t}=u_{xxx}+e^{-\alpha t}(c_{1}vv_{x}+c_{2}v),\\
v_{t}=vv_{x},
\end{array}
\label{sys5}
\end{equation}
\noindent
where $c_{1}$, $c_{2}$ and $\alpha$ are arbitrary constants. The recursion operator is
\begin{equation}
R=
\left(
\begin{array}{cc}
D^{2}+\alpha D^{-1} & e^{-\alpha t}\left(c_{1}v+c_{2}D^{-1}\right)  \\
0 & v
\end{array} \right).
\label{rec5}
\end{equation}
\newline {\bf iii)}
\begin{equation}
\begin{array}{c}
u_{t}=u_{xxx}+t^{-1/2}uu_{x},\\
v_{t}=\frac{1}{3}\left(t^{-1/2}uv_{x}+t^{-2/3}vv_{x}+t^{-5/6}u\right).
\end{array}
\label{sys6}
\end{equation}
\noindent This system admits the recursion operator $
R=
\left(\begin{array}{cc}
R^{0}_{0} & R^{0}_{1} \\
R^{1}_{0} & R^{1}_{1}
\end{array} \right)
$ with
\begin{equation}
\begin{array}{l}
R^{0}_{0}=tD^{2}+\frac{x}{3}+\frac{2}{3}t^{1/2}u\
+\frac{1}{3}(t^{1/2}u_{x}+\frac{1}{2})D^{-1}, \\
R^{0}_{1}=0, \\
R^{1}_{0}=\frac{1}{3}(t^{1/2}v_{x}+t^{1/6})D^{-1}, \\
R^{1}_{1}=\frac{1}{3}(t^{1/3}v+x).
\end{array}
\label{rec6}
\end{equation}

On the completeness of the above given non-autonomous systems we have the following proposition.

{\bf Proposition 3:} Systems (\ref{sys4}), (\ref{sys5}) and (\ref{sys6}) constitute all the nontrivial, integrable non-autonomous systems in $b(1)$ class. 

All other solutions of (\ref{inkos}) are trivial. Explicit solutions of (\ref{inkos}) with $b^{i}_{j}= b(1)$ prove propositions 2 and 3.

Systems related by  invertible transformations can be considered equivalent 
\cite{GKR,AG,KIN,FUCH1}. 
For this reason, we analyzed the above mentioned non-autonomous systems in regard to their transformability to autonomous systems via a certain generic class of  transformations. In the next proposition we present the result.  

{\bf Proposition 4:} None of the non-autonomous systems (\ref{sys4}), (\ref{sys5}) 
or (\ref{sys6}) is transformable to any autonomous system up to the invertible 
change of variables
\begin{eqnarray}
&x=\alpha(\tilde{t})\tilde{x}+\beta(\tilde{t}),\,\, t=\gamma(\tilde{t}),& \nonumber \\
&u(x,t)=\delta(\tilde{t})\tilde{u}(\tilde{x},\tilde{t})+\phi(\tilde{x},\tilde{t}),&\label{trns} \\
&v(x,t)=\rho(\tilde{t})\tilde{v}(\tilde{x},\tilde{t})+\psi(\tilde{x},\tilde{t}).& \nonumber
\end{eqnarray}
The $\alpha=0$ special case of system (\ref{sys5}) is the only exception. 
In this case (\ref{sys5}) is apparently autonomous.
This proposition can be verified by direct calculation for each of the mentioned systems.

\subsection*{b) Systems associated with $b^{i}_{j}=b(2)$}
~~~~Solutions of integrability conditions (\ref{inkos})
under $b^{i}_{j}=b(2)$ in (\ref{mtrcs}) give rise to the following nontrivial, integrable subclasses of systems.
\newline {\bf i)} 
\begin{equation}
\begin{array}{c}
u_{t}=v_{xxx}+vu_{x}+c_{1}uv_{x}+svv_{x}+yv,\\
v_{t}=vv_{x},
\end{array}
\label{sys8}
\end{equation}
\noindent
where $c_{1}$ is an arbitrary constant, is integrable if the undetermined functions 
$s=s(t)$ and $y=y(t)$ satisfy the differential constraint
\begin{eqnarray}
\frac{ds}{dt}+c_{1}y=0.
\end{eqnarray}
\noindent
These systems admit the recursion operator
\begin{equation}
R=
\left(
\begin{array}{cc}
(1-\frac{c_{1}}{2})v & D^{2}+c_{1}u+sv+\frac{c_{1}}{2}u_{x}D^{-1}  \\
0 & (1-\frac{c_{1}}{2})v+\frac{c_{1}}{2}v_{x}D^{-1}
\end{array} \right).
\label{rec8}
\end{equation}
\newline {\bf ii)} 
\begin{equation}
\begin{array}{c}
u_{t}=v_{xxx}+c_{1}(vu_{x}+2uv_{x})+svv_{x}+yv,\\
v_{t}=vv_{x},
\end{array}
\label{sys7}
\end{equation}
\noindent
where $c_{1}$ is an arbitrary constant, is integrable if the undetermined functions 
$s=s(t)$ and $y=y(t)$ satisfy the differential constraint
\begin{eqnarray}
\frac{ds}{dt}+(3c_{1}-1)y=0.
\label{difcon1}
\end{eqnarray}
\noindent
The recursion operator is 
\begin{equation}
R=
\left(
\begin{array}{cc}
0 & D^{2}+2c_{1}u+sv+c_{1}u_{x}D^{-1}  \\
0 & (1-c_{1})v+c_{1}v_{x}D^{-1}
\end{array} \right).
\label{rec7}
\end{equation}

In this subclass we observed that the $c_{1}=1$ particular case of (\ref{sys7}) with (\ref{difcon1}) admits two recursion operators $R(1)$ and $R(2)$. 
The first one being (\ref{rec7}) with $c_{1}=1$ and the second one is
\begin{equation}
R(2)=
\left(
\begin{array}{cc}
0 & tD^{2}+2tu+tsv+(tu_{x}-\frac{1}{2}s)D^{-1}  \\
0 & (tv_{x}+1)D^{-1}
\end{array} \right).
\label{rec7a}
\end{equation}
\newline {\bf iii)} 
\begin{equation}
\begin{array}{c}
u_{t}=v_{xxx}+t^{-2/3}(vu_{x}+c_{1}vv_{x})+yv,\\
v_{t}=t^{-2/3}vv_{x},
\end{array}
\label{sys9}
\end{equation}
\noindent
where $c_{1}$ is  an arbitrary constant, is integrable for any 
arbitrary function $y=y(t)$. The recursion operators of this system are
\begin{equation}
R(1)=
\left(
\begin{array}{cc}
\frac{x}{3}+t^{1/3}v & tD^{2}+c_{1}t^{1/3}v \\
0 & \frac{x}{3}+t^{1/3}v
\end{array} \right),\,\,\,
R(2)=
\left(
\begin{array}{cc}
0 & D^{-1}  \\
0 & 0
\end{array} \right).
\label{rec9}
\end{equation}
\noindent
The second operator $R(2)$ is not a proper recursion operator because of its nilpotent character. Nevertheless, $R(2)$ itself or any linear combination of it with $R(1)$ satisfies the conditions to be a recursion operator for (\ref{sys9}).
\newline {\bf iv)}
\begin{equation}
\begin{array}{c}
u_{t}=v_{xxx}+t^{\alpha}\left(uv_{x}+\frac{\alpha +1}{3\alpha +2}vu_{x}\
+svv_{x}\right)+yv, \\
v_{t}=\frac{\alpha +1}{3\alpha +2}t^{\alpha}vv_{x},
\end{array}
\label{sys10}
\end{equation}
\noindent
where $\alpha \neq -2/3$ is an arbitrary constant, 
is integrable if the undetermined 
functions $s=s(t)$ and $y=y(t)$ satisfy the differential constraint
\begin{eqnarray}
\frac{ds}{dt}+y=0.
\end{eqnarray}  
These systems admit the recursion operator
\begin{equation}
\begin{array}{l}
R^{0}_{0}=-\frac{\alpha}{2}\left( x +\frac{t^{\alpha +1}}{3\alpha +2}v \right),\\ 
R^{0}_{1}=tD^{2}+t^{\alpha +1}u+t^{\alpha +1}sv\
+\frac{1}{2}\left(t^{\alpha +1}u_{x}-(3\alpha +2)s\right)D^{-1},\\
R^{1}_{0}=0, \\
R^{1}_{1}=-\frac{\alpha}{2}\left( x +\frac{t^{\alpha +1}}{3\alpha +2}v \right)+\
\frac{1}{2}\left(t^{\alpha +1}v_{x}+3\alpha+2\right)D^{-1}.
\end{array} 
\label{rec10}
\end{equation}
\newline {\bf v)}
\begin{equation}
\begin{array}{c}
u_{t}=v_{xxx}+e^{t/\alpha}\left(vu_{x}+3uv_{x}+svv_{x}\right)+yv, \\
v_{t}=e^{t/\alpha}vv_{x},
\end{array}
\label{sys11}
\end{equation}
\noindent
where $\alpha \neq 0$ is an arbitrary constant, 
is integrable if the undetermined functions 
$s=s(t)$ and $y=y(t)$ satisfy the differential constraint
\begin{eqnarray}
\frac{ds}{dt}+3y=0.
\end{eqnarray} 
The recursion operator is
\begin{equation}
\begin{array}{l}
R^{0}_{0}=-\frac{1}{2}\left(x +\alpha e^{t/\alpha}v\right), \\ 
R^{0}_{1}=\alpha \left[D^{2}+3e^{t/\alpha}u+e^{t/\alpha}sv+\frac{1}{2}\
\left(3e^{t/\alpha}u_{x}-\frac{1}{\alpha} s\right)D^{-1}\right],\\
R^{1}_{0}=0, \\ 
R^{1}_{1}=-\frac{1}{2}\left(x +\alpha e^{t/\alpha}v\right)\
+\frac{3}{2}\left(\alpha e^{t/\alpha}v_{x}+1 \right)D^{-1}.
\end{array} 
\label{rec11}
\end{equation}
\newline {\bf vi)}
\begin{equation}
\begin{array}{c}
u_{t}=v_{xxx}+(uv)_{x}+svv_{x}+yv,\\
v_{t}=vv_{x},
\end{array}
\label{sys12}
\end{equation}
\noindent
is integrable for any arbitrary functions $s=s(t)$ and $y=y(t)$. These systems admit 
the recursion operators $R(1)$, $R(2)$ and  $R(3)$ where 
\begin{equation}
\begin{array}{l}
R(1)^{0}_{0}=-v_{x}D^{-1}, \\
R(1)^{0}_{1}=D^{2}+2u-2(\int{ydt})v+\left[u_{x}+(\int{ydt})v_{x}\right]D^{-1},~~~~~~~~~~~\\
R(1)^{1}_{0}=0,\\ 
R(1)^{1}_{1}=v_{x}D^{-1},\\
\end{array} 
\end{equation}
\begin{equation}
\begin{array}{l}
R(2)^{0}_{0}=-x-tv-2(tv_{x}+1)D^{-1}, \\
R(2)^{0}_{1}=2tu+2(\int{(s-ty)dt})v\\
~~~~~~~~~~~+\left[tu_{x}+\left(\int{(s+2ty+3\int{ydt})dt} \right)v_{x}\
+3\int{ydt}\right]D^{-1}, \\
R(2)^{1}_{0}=0,  \\
R(2)^{1}_{1}=-x-tv+(tv_{x}+1)D^{-1},\\
\end{array} 
\label{rec12}
\end{equation}
\noindent
\begin{equation}
R(3)=
\left(
\begin{array}{cc}
0 &  x+tv \\
0 & 0
\end{array} \right).
\end{equation}
Obviously, $R(3)$ is nilpotent and thus it is not a proper recursion operator.
\newline {\bf vii)} 
\begin{equation}
\begin{array}{c}
u_{t}=v_{xxx}+\frac{1}{\sqrt{t}}(uv)_{x}+svv_{x}+yv,\\
v_{t}=\frac{1}{\sqrt{t}}vv_{x},
\end{array}
\label{sys13}
\end{equation}
\noindent
is integrable for any arbitrary functions $s=s(t)$ and $y=y(t)$. These systems admit 
the recursion operators $R(1)$, $R(2)$ and $R(3)$ where 
\begin{equation}
\begin{array}{l}
R(1)^{0}_{0}=x+2\sqrt{t}v+\left(2\sqrt{t}v_{x}+1\right)D^{-1}, \\
R(1)^{0}_{1}=2tD^{2}-\left[\left(\int{(2\sqrt{t}y\
+\frac{1}{\sqrt{t}}\int{ydt})dt}\right)v_{x}\
+\int{ydt}\right]D^{-1}, \\
R(1)^{1}_{0}=0,\\ 
R(1)^{1}_{1}=x+2\sqrt{t}v,
\end{array} 
\end{equation}
\begin{equation}
\begin{array}{l}
R(2)^{0}_{0}=x+2\sqrt{t}v, \\
R(2)^{0}_{1}=4tD^{2}+4\sqrt{t}u+2\left(\int{(s-2\sqrt{t}y)dt}\right)v\\
~~~~~~~~~~~+\left[2\sqrt{t}u_{x}+\left(\int{(s+\frac{1}{\sqrt{t}}\int{ydt})dt}\right)v_{x}\
+\int{ydt}\right]D^{-1}, \\
R(2)^{1}_{0}=0,  \\
R(2)^{1}_{1}=x+2\sqrt{t}v+\left(2\sqrt{t}v_{x}+1\right)D^{-1},\\
\end{array}
\label{rec13} 
\end{equation}
\begin{equation}
R(3)=
\left(
\begin{array}{cc}
0 &  x+2\sqrt{t}v \\
0 & 0
\end{array} \right).
\end{equation}
Again we obtain an improper recursion operator $R(3)$.

There are other solutions to integrability conditions (\ref{inkos}) in $b(2)$ class as well. However, those solutions correspond to trivial systems or recursion operators. For this reason we have the following proposition.

{\bf Proposition 4:} The subclasses (\ref{sys7})-(\ref{sys13})  constitute all 
the nontrivial subclasses in $b(2)$ class.

 About the transformability of non-autonomous systems in the above subclasses, we have the following proposition.

{\bf Proposition 5:} Non-autonomous systems in each of the subclasses 
(\ref{sys7})-(\ref{sys13}) of $b(2)$ class are not transformable to any autonomous system 
through the transformations (\ref{trns}).

It has been recently observed that some  recursion operators which are called weak, do not always generate symmetry hierarchies correctly \cite{SW}. Some of the operators we found in this work are of this type. The source of this weakness and possible solutions which relies on the found weak recursion operators are extensively investigated in \cite{GKR2}.  

{\bf Acknowledgements}

The author would like to thank  professors M. G{\"u}rses and A. Karasu  for their valuable comments and discussions and also Dr. S.Y. Sakovich for his reading and suggestions on the manuscript. This work is partially supported by the Scientific and Technical Research Council of Turkey (T{\"U}B{\.I}TAK).

\end{document}